\documentstyle[aps,epsf,epsfig,multicol]{revtex}

\begin{document}

\title{Localization properties of groups of eigenstates in chaotic systems}
\author{D. A. Wisniacki$^{1,2}$, F. Borondo$^{1}$, E. Vergini$^{2}$,
and R. M. Benito$^{3}$ }
\address{$^1$ Departamento de Qu\'{\i}mica C--IX,
 Universidad Aut\'{o}noma de Madrid,
 Cantoblanco, 28049--Madrid (Spain).}
\address{$^2$ Departamento de F\'{\i}sica,
 Comisi\'on Nacional de Energ\'{\i}a At\'omica.
 Av.\ del Libertador 8250, 1429 Buenos Aires (Argentina).}
\address{$^3$ Departamento de F\'{\i}sica y Mec\'{a}nica,
 E.T.S.I.\ Agr\'{o}nomos, Universidad Polit\'{e}cnica de Madrid,
 28009--Madrid (Spain).}
\date{Received \today}
\maketitle

\begin{abstract}
In this paper we study in detail the localized wave functions defined
in Phys.\ Rev.\ Lett.\ {\bf 76}, 1613 (1994), in connection with the
scarring effect of unstable periodic orbits in highly chaotic
Hamiltonian system.
These functions appear highly localized
not only along periodic orbits but also on the associated manifolds.
Moreover, they show in phase space the hyperbolic structure in the
vicinity of the orbit, something which translates in configuration
space into the structure induced by the corresponding self--focal points.
On the other hand, the quantum dynamics of these functions are
also studied.
Our results indicate that the probability density first evolves
along the unstable manifold emanating from the periodic orbit,
and localizes temporarily afterwards on only a few, short related
periodic orbits.
We believe that this type of studies can provide some keys to
disentangle the complexity associated to the quantum mechanics of these
kind of systems,
which permits the construction of a simple explanation in terms of the
dynamics of a few classical structures.
\end{abstract}

\pacs{PACS numbers: 05.45.+b, 03.65.Sq, 03.20.+i}

\begin{multicols}{2}
\section{Introduction}
  \label{sec:intro}

The investigation of the quantum manifestations of classical chaos is
at present a very active field of research \cite{gut}.
The relationship between quantum dynamics and classical invariants,
mainly periodic orbits (PO) and their associated manifolds,
is not satisfactorily understood yet \cite{qchaos}.
In this respect, an important achivement would be to disentangle the
time evolution of a quantum system whose classical analog is chaotic,
explaining this dynamics in terms of classical structures.

Fifteen years ago, a seminal step was done by Heller \cite{hel84},
who considered the evolution of a wave packet localized on an unstable PO.
He found that the dynamics in the short time limit are controlled by the
motion around the orbit and its Lyapunov exponent and is well understood
in terms of semiclassical expressions \cite{hel84,houches}.

For longer times the dynamical mixing induced by the chaotic behavior
of the motion makes the situation more complicated, and one single orbit
and the linearized dynamics around it is not enough to account for the
evolution of the wave packet.
Some time later, Tomsovic and Heller \cite{tom} showed that even for
those long times, for which classical fine structure had developed on
a scale much smaller than $\hbar$, the semiclassical propagation of
the packet can be carried out with good precission, by computing the
corresponding correlation function, $C_{\rm scl}(t)$, as a sum of
contributions of the homoclinic excursions of the PO.
This procedure has nevertheless some drawbacks, since for example the
number of orbits that is necessary to include in the calculation to
obtain converged results grows dramatically with time.

A different approach, that maintains an interpretation of wave mechanics
based on the more simple classical objects represented by POs has been
presented elsewhere \cite{ver,PRE}.
In this theory \cite{ver} all quantum information of a bounded chaotic
Hamiltonian system can be obtained using a number of POs that grows only
linearly with the Heisenberg time.

Another point worth considering is the importance and/or convenience
of using wave functions which are ``dynamically adapted'' in this
type of studies .
By ``dynamically adapted'' \cite{sch} we mean structures that contain
information not only about the PO, but also about the linearized
dynamics around the orbit.
When considered in phase space, these structures appear no only
localized over the PO, but are also
spread over the corresponding manifolds.

The construction of functions of this kind can be carried out using
different approaches.
In Ref.~\onlinecite{pol} the necessary information is obtained by
Fourier transforming the short--time exact quantum dynamics, so that wave
functions averaged over the PO path are obtained.
Kaplan and Heller \cite{kap} used a sum of coherent Gaussian wave
packets centered along the PO to obtain the desired function.
Finally, Vergini and Carlo \cite{ver2} constructed semiclassically
resonances along POs with minimum energy dispersion.

In spite of the differences existing among these functions, they are
all suitable to efficiently investigate the quantum manifestations
of the classical phase space complexity in classically chaotic systems.

The purpose of this paper is threefold.
In the first place, we focus on the analysis of the topological
characteristics (i.e.\ distribution of quantum probability)
in phase space of this type of wave functions highly localized on an
unstable PO.
In the second place, special attention will be paid to the relationship
existing between the quantum dynamics of this function and that of other
classical POs of the system located in their vicinity, in order to
explore the existence of dynamical connections and active interplay
between them.
Finally, another interesting aspect of these functions is their revivals,
that take place in some cases for a surprisingly large number of cycles
without showing any visible sign of dispersion \cite{revivals}.
This fact, which is particularly surprising in view of the highly
chaotic character of the problem that will be considered,
is due to the special dynamical properties of these functions.
Summarizing, it should be stressed that the most important result of
our work is that, proceeding in this way, only a few short POs are
necessary to adequately explain the quantum dynamics of this type of
systems.

For our study we will use localized functions as defined in
Ref.~\onlinecite{pol}, applying them to the study of the dynamics
in the Bunimovich stadium billiard \cite{bun}, a system which is
known to behave ergodically from the classical point of view.

The organization of the paper is as follows.
In Sect.~\ref{sec:tool} we describe in detail how the method of
Ref.~\onlinecite{pol} can be used to construct the desired wave functions,
highly localized on a given unstable PO and their associated manifolds.
Numerical results concerning the topological characteristics, both in
configuration and phase space, of these functions and the corresponding
time evolution will be presented and analyzed in Sects.~\ref{sec:topo}
and \ref{sec:dyn}, respectively.
Finally, the main conclusions derived from our work are presented in
Sect.~\ref{sec:final}.
\section{Constructing Localized Wave Functions}
  \label{sec:tool}

In this section we briefly describe the method \cite{pol} that will
be used to construct states initially located on a given PO and the
corresponding  manifolds.
The procedure has been previously applied to the analysis of the scarring
effect of unstable POs in chaotic systems \cite{prev}, as well as
to the study of the quantum dynamics that takes place beyond the first
recurrence of the corresponding motion \cite{PRE}.
As it will be shown, these initial states can be used as an important
tool to disentangle the complexity of the full quantum dynamics taking
place in the area of influence of a given PO, making use of the
associated classical mechanics as a guide in the process.

Initially, the method considers a wave function, $|\phi(0)\rangle$,
well localized in phase space in the vicinity of a given PO.
The corresponding time evolution can be followed by means of the
associated autocorrelation function,
%
\begin{equation}
  C(t) =  \langle \phi(0) | \phi(t) \rangle =
          \langle \phi(0) | e^{-i{\hat H}t} | \phi(0) \rangle.
 \label{eq:1}
\end{equation}
Recurrences in $C(t)$ are known
to determine the low resolution structure of the corresponding spectra
\cite{houches},
%
\begin{equation}
  I_{T}(E) = \frac{1}{2\pi} \int_{-\infty}^\infty \; dt \; W_{T}(t) \; C(t) \; e^{iEt} \:,
 \label{eq:2}
\end{equation}
and are the origin of the scarring effect, as first discussed
in \cite{hel84}. $W_T(t)$ is a  window function to take into account
possible different resolutions in $I_T(E)$.

A widely spread choice for $|\phi(0)\rangle$ is a minimun uncertainty
harmonic oscillator coherent state ($\hbar$ is set equal to one
throughout this paper)
%
\begin{eqnarray}
  G_{{\bf r}^0,{\bf P}^0}({\bf r}) & = &
  \prod_j \left(\frac{1}{\pi\sigma_j^2}\right)^{1/4}
  \exp\left[\frac{-(r_j-r_j^0)^2}{2\sigma_j^2} \right]
                                            \nonumber \\
   & & \times \exp\left[iP_j^0 (r_j-r_j^0) \right],
  \label{eq:3}
\end{eqnarray}
where (${\bf r}^0, {\bf P}^0$) represents the coordinates and conjugate
momenta of a suitable phase space point along the selected PO.

For bounded systems, the evolution of this packet can be
followed quite conveniently by projection on a complete set of
eigenfunctions, $|n\rangle$, of the Hamiltonian, ${\hat H}$,
%
\begin{equation}
  |\phi(t)\rangle = e^{-i{\hat H}t} | \phi(0) \rangle =
   \sum_n |n \rangle \langle n|\phi(0)\rangle \; e^{-i E_n t},
 \label{eq:4}
\end{equation}
where $\{E_n\}$ is the corresponding set of eigenvalues.
In our case the billiard eigenstates have been computed using the
scaling method \cite{sca}.

Regarding the smoothing function $W_T(t)$ one has many choices;
i.e.\ ``hat'', exponential or Gaussian functions, which render
$\sin x /x$, Lorentzian or Gaussian--type line envelopes,
respectively.
In our case we use the Gaussian decay:
%
\begin{equation}
   W_T(t) = \frac{e^{-t^2/2T^2}}{T},
 \label{eq:5}
\end{equation}
for which the spectrum takes the form
%
\begin{equation}
  I_T(E) = \frac{1}{(2\pi)^{1/2}} \sum_n |\langle n|\phi(0)\rangle|^2 \;
      e^{-T^2(E-E_n)^2/2}
 \label{eq:6}
\end{equation}

The center of the initial wave packet defined in Eq.~(\ref{eq:3}) will
follow for some time the classical PO on which it was launched,
with the dispersion rate evolving according to the Lyapunov exponent
of the orbit \cite{houches}.

In a second step \cite{pol}, a wave function highly scarred by the PO
can be constructed very efficiently by considering the average of
$|\phi(t)\rangle$ along the dynamics generated by the PO during a
given amount of time, usually taken of the order of the PO period.
Let us remark the importance of this last requirement, since if the
packet is allowed to evolve for a very long time, so that due to the
chaotic character of the dynamics it has the opportunity to sample
all the available phase space, just the (complicated) eigenfunctions of
the system will be obtained.
The corresponding expression for the localized wave function corresponding
to an energy $E_0$ (usually taken as the center of one of the bands
appearing in the low resolution spectrum generated by $|\phi(0)\rangle$)
reads as follows:
%
\begin{eqnarray}
  |\psi^{E_0}\rangle & = & \frac{1}{2\pi} \; \int_{-\infty}^\infty \;
    dt \; e^{iE_0t} W_T(t) |\phi(t)\rangle
                         \nonumber \\
    & = & \frac{1}{(2\pi)^{1/2}} \sum_n
    | n \rangle \langle n | \phi(0) \rangle \; e^{-T^2(E_0-E_n)^2/2}
 \label{eq:7}
\end{eqnarray}

Phase space representations for the wave functions described in this
section can be constructed in a number of different ways.
In this paper we will follow the procedure described in
Ref.\onlinecite{tua}, which relies on the use of normal derivatives
of the wave functions evaluated at the boundary of the billiard.
On this boundary, Birkhoff coordinates \cite{arn}, ($q,p$), are used
to define both classical and quantal Poincar\'e surfaces of section,
such that $q$, is the arc length
coordinate, and $p={\bf p}\cdot{\bf{\hat t}}/|{\bf p}|$ is the fraction
of tangential momentum.
The coherent states necessary to construct this representation (Husimi
functions \cite{hus}) are then defined as:
%
\begin{eqnarray}
  G_{q^0 \; p^0}(q)  & = & \left( \frac{1}{\pi\sigma^2} \right)^{1/4}
  \exp \left[-\frac{1}{2\sigma^2} (q-q^0)^2 \right]
                                                       \nonumber \\
    &  & \times \exp \; [ip^0 (q-q^0)].
 \label{eq:8}
\end{eqnarray}
This expression corresponds to a boundary wave packet at point
($q^0,p^0$) in the surface of seccion;
representing a bounce off a given boundary point with a specific
tangential momentum. Then, for a wavefunction with normal derivative
on the boundary $\psi(q)$ (extended periodically to the real line),
the corresponding Husimi function is given by

\begin{equation}
  H(q^0,p^0) = \left| \langle G_{q^0 \; p^0}|\psi \rangle\right|^2
 \label{eq:9}
\end{equation}
\section{Topological Characteristics of the Localized Wave Functions}
  \label{sec:topo}

The system that we have chosen to study is a particle of mass 1/2 moving
in a desymmetrized stadium billiard with Newman boundary conditions on
the symmetry axis (only even--even parity wave functions will then be
considered).
The radius $r$ is taken equal to unity and the enclosed area is 1+$\pi$/4.
This system is known to exhibit a great degree of chaoticity,
both from the classical and quantal points of view.

In this case, and due to the symmetry of the problem,
symmetry adapted initial wave functions should be used;
the corresponding expression being given by
%
\begin{eqnarray}
  \langle x,y|\phi(0)\rangle & = &
   G_{x^0,y^0,P_x^0,P_y^0} +  \; G_{-x^0,y^0,-P_x^0,P_y^0}
                                                     \nonumber \\
   & + &  \; G_{x^0,-y^0,P_x^0,-P_y^0} +  
   \; G_{-x^0,-y^0,-P_x^0,-P_y^0} \nonumber \\ 
   & + &  \; c.c.
 \label{eq:10}
\end{eqnarray}
is obtained by imposing Newman
boundary condition at the symmetry axis.

\begin{figure}
\centering \leavevmode
\epsfxsize=4cm
\center{\epsfig{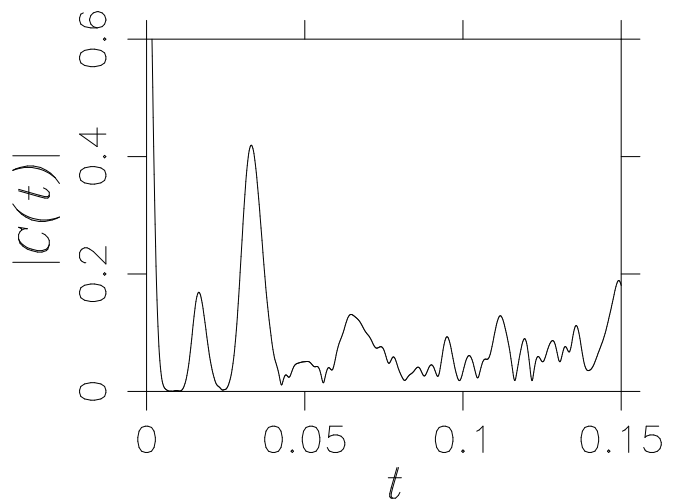}}
\vspace{0.25cm}
\caption{Modulus of the autocorrelation function corresponding to a
   symmetrized wave packet [Eq. (\protect\ref{eq:9})] initially centered on
   the horizontal periodic orbit of the desymmetrized stadium billiard
   with Newman boundary conditions on the axis at an energy $E$=3600.
   The radius is taken equal to 1, the enclosed area 1+$\pi$/4,
   and the mass of the particle 1/2.}
 \label{fig:1}
\end{figure}

We consider the dynamics dominated by the horizontal PO,
running along the $x$ axis with $y=0$.
Accordingly, we use for our study a symmetry adapted Gaussian wave packet
[Eq. (\ref{eq:10})] with centers defined by the phase space point
$(x^0,y^0,P_x^0,P_y^0)=(1,0,k,0)$ and width
$\sigma_x = \sigma_y = 1.603/k^{1/2}$, where $k$ is the usual wave number.
The corresponding autocorrelation function for a packet with an energy
at the center of $E$=3600, for which the period of the horizontal PO
is equal to $T_0$=1/30, is shown in Fig.~\ref{fig:1}.
As can be seen, two main peaks exist at $t \simeq 0.016$ and 0.033
respectively, being the first one substantially lower than the second.
After that, the correlation function becomes very complicated.
The relative intensity of the two peaks can be easily explained if one
takes into account that the initial coherent state that we are using
is the sum of two packets, launched with opposite values of the momentum.
The two packets first separate, one being scattered by the vertical
axis and the other by the circular hard wall of the billiard, so that
when they first return to the initial point, at a time roughly equal
to $T_0/2$, they have picked up different phases.
This gives rise to the first (small) recurrence peak in $|C(t)|$.
Afterwards, the two components continue moving, colliding with the
circular and vertical walls, respectively.
In this way, when some time later, at $t \simeq T_0$, they return again
close to the initial point, both arrive with approximately the same
collisional phase, this giving rise to the second, more pronounced peak
in the autocorrelation function.

\begin{figure}[t]
\centering \leavevmode
\epsfxsize=4cm
\center{\epsfig{file=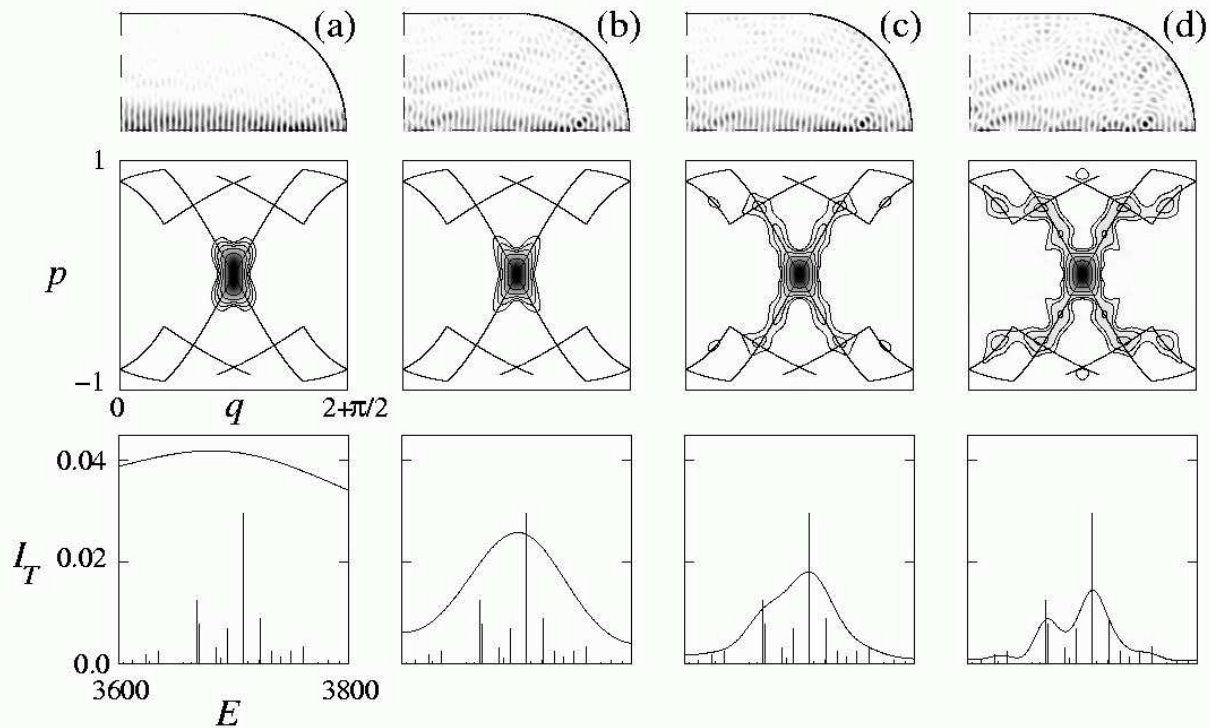, ,width=10cm,angle=-90}}
\vspace{0.25cm}
 \caption{
  (Top) Localized (squared) wave functions calculated using
        Eq.~(\protect\ref{eq:7}),
  (middle) Husimi based quantum Poincar\'e surface of section, and
  (bottom) spectra ($I_\infty$ as a stick spectrum and $I_T$ in
  solid line) for different values of the smoothing time, $T$:
  (a) 0.01, (b) 0.03, (c) 0.06, and (c) 0.1 at $E_0$=3700, corresponding
  to the autocorrelation function of Fig.~\protect\ref{fig:1}.
  The invariant manifolds of the horizontal periodic orbit have also
  been drawn superimposed in the middle pannel.}
 \label{fig:2}
\end{figure}

Let us consider next the localization of the wave functions
obtained with the method described in the previous section
[see Eq.~(\ref{eq:7})] for different values of the smoothing time.
We choose $T$=0.01, 0.03, 0.06, and 0.1, in order
to cover a range of values smaller and greater than $T_0$.
The results are shown in Fig.~\ref{fig:2}, where we present the
corresponding squared wave functions, Husimi based quantum sufaces of
sections, and the associated portions of the spectra generated from
$|\phi(0)\rangle$.
Three comments are in order.
\end{multicols}
\begin{multicols}{2}
In the first place, the wave functions in configuration space (top tier)
appear localized along the PO, presenting a number of nodes, $m$,
in agrement with the value obtained from the Bohr--Sommerfeld
quantization condition
%
\begin{equation}
  k=\frac{2\pi}{L} \left( m+\frac{\nu}{4} \right)
 \label{eq:11}
\end{equation}
when applied to the spectral peaks (bottom tier).
For the horizontal PO, $L=4$ and the Maslov index $\nu$=3.

In the second place, and much more importantly, in the localized wave
functions it is clearly observable the distorsion in the structure
induced by the self focal point.
A discussion of this effect can be found in Ref.~\onlinecite{houches}.
This is a very important point, and is a consequence of the fact that
these functions have been defined in such a way--averaging over the
motion along the orbit--that they are not only located in the right
position of configuration space, but contain also information
on the dynamics of the system in the vicinity of the PO.
This is more clearly seen when examining the phase space picture of these
functions, i.e.\ middle tier of Fig.~\ref{fig:2}.
There, it is observed that the probability density do not just localize
over the fixed point corresponding to the scarring orbit, but it is also
spread significantly along their manifolds.

Finally, this structure over the manifolds changes as the value of $T$
is increased, giving the original packet the opportunity to explore
more and more of the dynamics induced by the PO.
Indeed, we see in Fig.~\ref{fig:2} that for $T$=0.01, well before the
first peak in $|C(t)|$, the quantum surface of section appear essentially
localized on the fixed phase point corresponding to the horizontal PO,
with just some probability sticking out of this region along the
manifolds (also plotted in the figure).
For $T$=0.03 this latter effect is more pronounced, and the pattern
along the linearized part of the stable and unstable manifolds is
fully developed.
At the same time, the corresponding low resolution spectral envelopes
$I_T(E)$ (continuous line in the bottom tier) define, with increasing
precission, the clump of eigenstates contributing to this scar \cite{pol}.
Past this time, i.e.\ for $T$=0.06, well beyond the first (true)
recurrence of the horizontal PO, the original packet has had the
opportunity to enter into the truly nonlinear part of the periodic
trajectory, exploring regions of phase space far apart from the fixed
point, where the dynamical flux gets more complicated.
However, we see that the averaged wave function is well confined
on the manifolds, even following the sharp kinks that they present.
This process continues for $T$=0.1, value at which the localized wave
function appears covering practically all these invariant structures
of phase space.
In this case the resolution in the spectra is higher,
and the corresponding spectral envelope define several bands,
four in our case.
These higher resolution bands \cite{PRE} correspond to the interaction
with other POs of longer periods, as will be discussed in the next
section. The signatures of these longer POs are seen in the
configuration space plots.
\vspace{0.5cm}
%
\begin{figure}[t]
\centering \leavevmode
\center{\epsfig{file=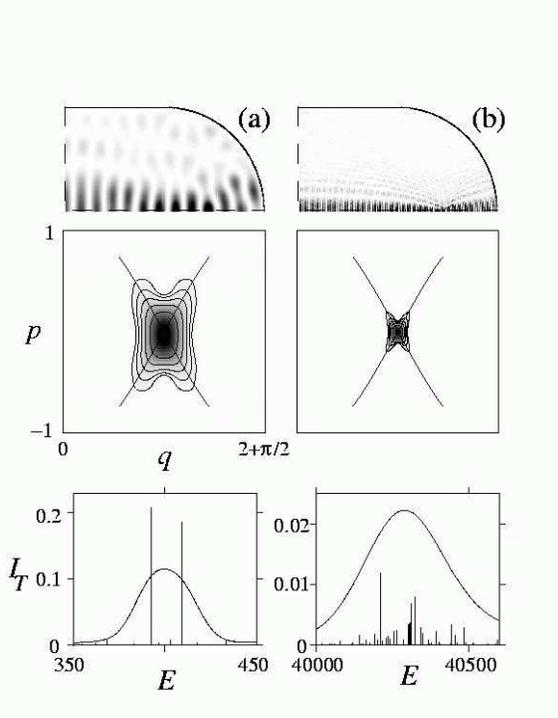, ,width=6.75cm,angle=0}}
\vspace{0.5cm}
  \caption{
  (Top) Localized (squared) wave functions calculated using
        Eq.~(\protect\ref{eq:7}),
  (middle) Husimi based quantum Poincar\'e surface of section, and
  (bottom) spectra ($I_\infty$ as a stick spectrum and $I_T$ in
  solid line) for:
  (a) $E_0$=400 and $T$=0.1, and (b) $E_0$=40267 and $T$=0.01.
  The invariant manifolds of the horizontal periodic orbit have also
  been drawn superimposed in the middle pannel.}
 \label{fig:3}
\end{figure}
\section{Dynamics of the Localized Wave Functions}
  \label{sec:dyn}

Another aspect of this problem which is worth considering is how the
process described above is affected when the transition to the
semiclassical limit is considered.
For this purpose we have repeated the previous calculations,
launching the packet from the same point as before but with different
values of the energy. The results are shown in Fig.~\ref{fig:3}.
In it we present the localized wave functions, Husimi based quantum
Poincar\'e surfaces of section, and spectra for $E_0$=400 and $T$=0.1
[column (a)], and also for $E_0$=40267 and $T$=0.01 [column (b)].
These values of the smoothing times correspond to the periods of the
horizontal PO at the selected energies, so that there are to be compared
to those in Fig.~\ref{fig:2} (b).
As can be seen, the structure exhibited by these functions is essentially
the same, if one takes into account the fact that all length features
scale as $k^{-1/2}$.
This scaling is the origin of the better definition of the focalization
effect existing in the configuration space wave functions at higher
energies.

Using Gutzwiller trace formulae, similar smoothed wave function has
recently been calculated \cite{diego}, also showing that the invariant
hyperbolic classical structure is contained in quantum mechanics.
In this case no dynamical information was used, with the result that many
more (of the order of 10000) states were needed in the summation in order
to get the same kind of localization that we are showing in this paper.

In this section we investigate the dynamics of the localized wave
functions described in the previous section, with the purpose
of disentangling the complexity of the quantum dynamics in the problem
that we are considering.
As stated in the Introduction, our aim is to provide an explanation
of the quantum dynamics of highly chaotic systems in terms of classical
invariant structures.
\begin{figure}
\centering \leavevmode
\epsfxsize=4cm
\center{\epsfig{file=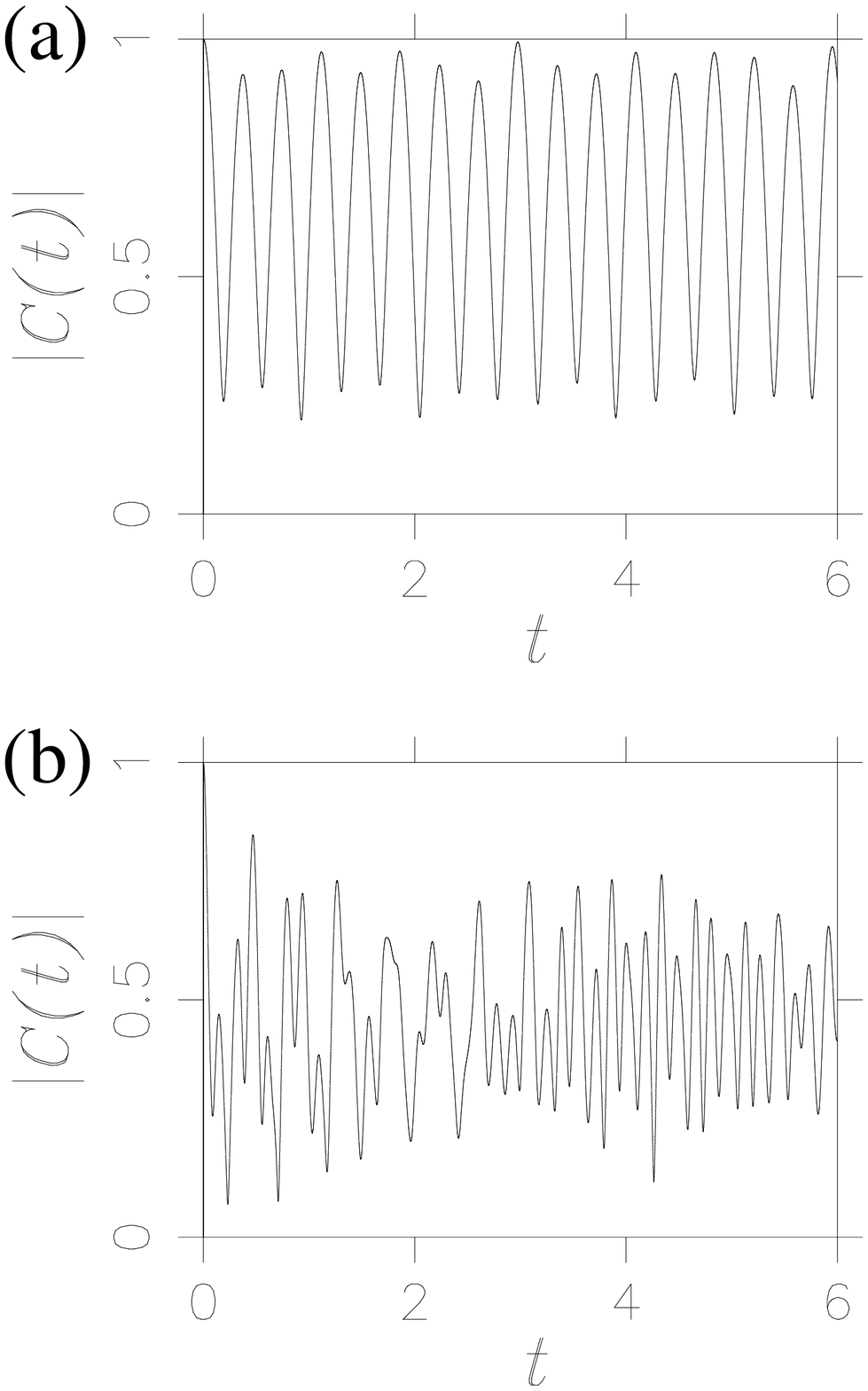, ,width=6.cm,angle=0}}
  \caption{Modulus of the autocorrelation function corresponding to
   localized wave functions calculated for:
   (a) $E_0$=400 and $T$=0.1, and
   (b) $E_0$=3700 and $T$=0.03.}
 \label{fig:4}
\end{figure}

In Fig.~\ref{fig:4} we present the autocorrelation function calculated
using the (non--stationary) localized wave functions [Eq. (\ref{eq:7})]
of Fig.~\ref{fig:3} (a) and Fig.~\ref{fig:2} (b).
The results of part (a), corresponding to $E_0$=400 and $T$=0.1
, show a series of recurrences
at time intervals approximately equal to 0.38.
In each of them, practically all the initial probability is recovered,
and the process shows no evidence of dispersion up to the times
considered in our calculations.
\end{multicols}

\begin{figure}
\centering \leavevmode
\epsfxsize=4cm
\center{\epsfig{file=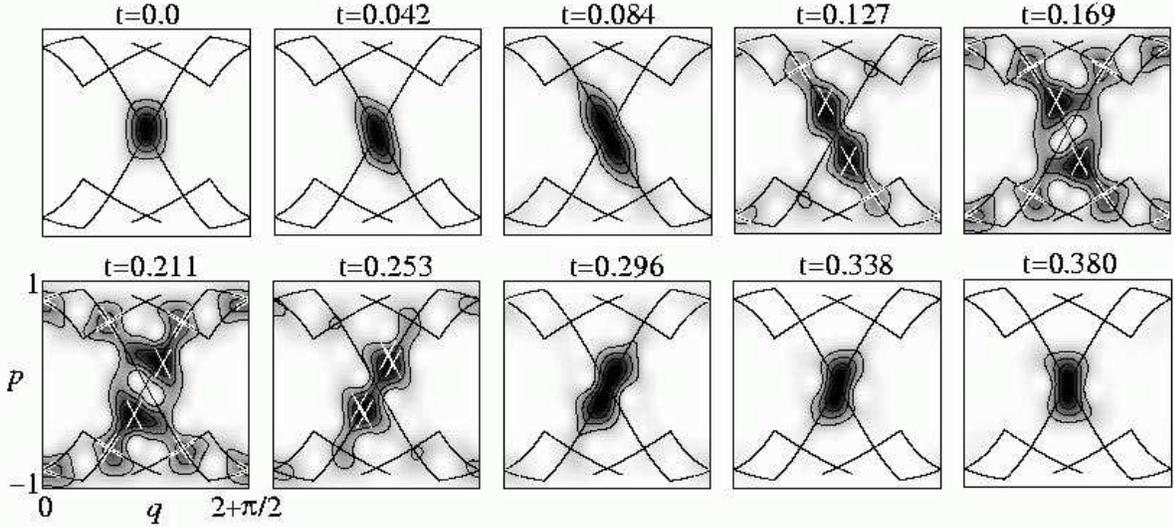, ,width=7.5 cm,angle=-90}}
\vspace{0.85cm}
\caption{Snapshots of the dynamical evolution in phase space of the
localized wave function corresponding to Fig.~\protect\ref{fig:3}
(a). The manifolds of the horizontal orbit are plotted in solid line.
The linearized manifolds
of the orbits (1), (2), (3) and (4) of Fig.~\protect\ref{fig:6} are included in
white lines (in the case that the Husimi function is localized on them). }
 \label{fig:5}
\end{figure}

\begin{multicols}{2}
This phenomenon has already been discussed in Ref.~\onlinecite{revivals},
where Gaussian wave packets ``stretched'' along the horizontal PO were
used, and necessary conditions for the existence of revivals were given.
As indicated before, in our case we use initial functions that are
better adapted to the dynamics of the orbit, thus showing an improved
revival behaviour.
In this respect, it is important to note that, opossed to which it was
found in \cite{revivals}, our functions show revivals, in this wavelength
regime, for an ample range of energy values and also for other POs.

\begin{figure}
\centering \leavevmode
\epsfxsize=4cm
\center{\epsfig{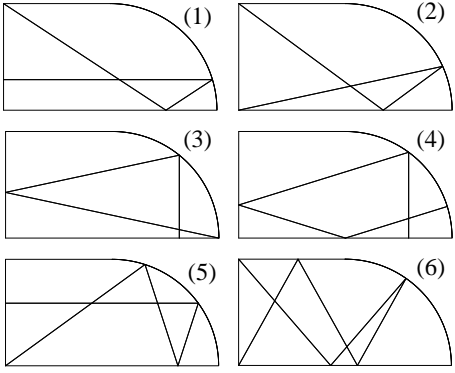}}
\vspace{0.8cm}
\caption{Some periodic orbits of the desymmetrized stadium billiard
related to the horizontal one. See text for details.}
\label{fig:6}
\end{figure}

\begin{figure}
\centering \leavevmode
\center{\epsfig{file=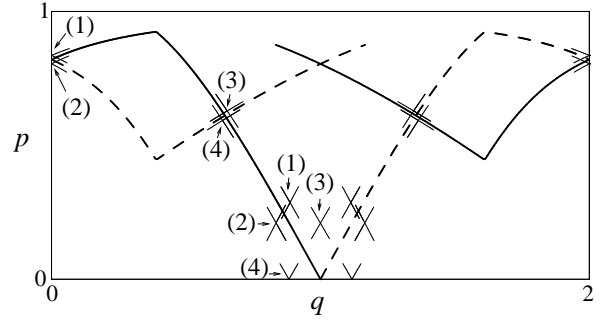, ,width=4.5cm,angle=-90}}
\vspace{0.8cm}
\caption{Fixed points and associated manifolds of some periodic
orbits in Fig.~\protect\ref{fig:6}.The unstable (thick full line)
and stable (thick dashed line) manifolds of the horizontal periodic
orbit are also represented.}
\label{fig:7}
\end{figure}

From the quantum mechanical point of view these recurrences are easy
to interpret, since they are the result of the beating frequency
originated from the two (almost exclusive) eigenstate contribution to
the initial packet (see spectrum in the bottom part of Fig.~\ref{fig:3}).
On the other hand, at the classical level, the phenomenon is
more difficult to understand.
To further investigate these recurrences from this point of view,
we show in Fig.~\ref{fig:5} a series of snapshots in phase space
covering the period of time elapsed up to the first recurrence of
$|\psi^{E_0}\rangle$.
Notice that due to the symmetry of the figure we have only plotted one
half of the complete phase space of the billiard.
To help interpreting this figure, the position of the fixed points and
associated manifolds, corresponding to some short POs related to the
horizontal one, have also been included.
These orbits are presented in Fig.~\ref{fig:6}.
For the sake of clarity, the labelling of the POs was not included in
Fig.~\ref{fig:5}, but it can be found in Fig.~\ref{fig:7}.

In the first three images of Fig.~\ref{fig:5} we see how the original
probability density distribution elongates, stretching along the unstable
manifolds of the horizontal PO and contracting along the stable ones.
Afterwards, the process continues, populating the regions around the
fixed points corresponding to orbits (3),(4) and (1),(2) \cite{footnote},
so that at $t$=0.127--0.169 a noticeably accumulation of probability
around these points has taken place.
Notice also the effect of the flux along the stable manifolds of
these POs, which at $t$=0.127 is narrowing the elongated shape of the
distribution along some specific lines, and at $t$=0.169 has split the
packet in several pieces.
At $t$=0.211 this process reaches its maximum, and the probability
remaining in the initial region is practically zero.
After this point, the evolution continues in a similar way with the
probability returning gradually to the vicinity of the initial point,
although obviously this time following a path going along the stable
manifolds.
This second half of the dynamics takes place following the same series
of intermediate steps.

\begin{figure}
\centering \leavevmode
\center{\epsfig{file=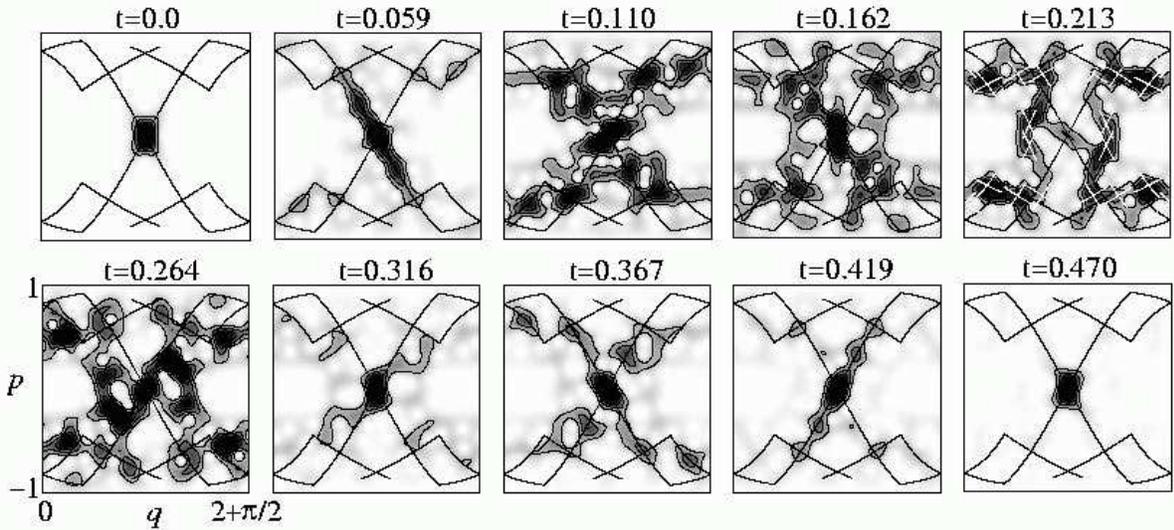, ,width=7.5 cm,angle=-90}}
\vspace{0.85cm}
\caption{Snapshots of the dynamical evolution in phase space of the
localized wave function corresponding to Fig.~\protect\ref{fig:2}
(b). The manifolds of the horizontal orbit are plotted in solid line.
The linearized manifolds
of the orbits (5) and (6) of Fig.~\protect\ref{fig:6} are included in
white lines at $t=0.213$. }
\label{fig:8}
\end{figure}

In Fig.~\ref{fig:4} (b) the autocorrelation function for the initial
state of Fig.~\ref{fig:2} (b) is shown.
In this case the recurrence pattern is much more complex,
but it is nevertheless noticeable the existence of a fairly good number
of recurrences in which the overlap with the initial distribution is
greater than 50\%.
The difference with the example presented in the part (a) of the figure
is clear. In this case, the initial wave function is formed by a
substantially greater number of contributions [see the associated
spectrum at the bottom of Fig.~\ref{fig:2} (b)], and then the
probability that all off them get in phase is much smaller.
%
The corresponding evolution in phase space is presented, up to the first
big recurrence at $t \simeq 0.47$, in Fig.~\ref{fig:8}.
At this higher energy we see that the mechanism for the dispersion of
the probability density is practically the same,
although there exists certain differences worth commenting.
In the first place, the time scale for the dispersion is obviously
faster, i.e.\ compare the extent of the evolution in the first snapshot
at $t$=0.059 with the third one at $t$=0.127 of Fig.~\ref{fig:5}.
In the second place, the probability distribution appears in this case
much more localized on the participating invariant classical structures
for all values of time considered.

\end{multicols}
\vspace{1.5cm}
\begin{multicols}{2}
Finally, at certain values of $t$,
the probability show peaks which localize on new POs, not important
for the dynamical evolution at $E_0$=400 shown in Fig.~\ref{fig:5}.
For example, at $t=0.213$ the probability density localize on
orbits (5) and (6) of Fig.~\ref{fig:6}.
This effect is not unexpected, since it is reasonable to admit that
as we increase the energy, going towards the semiclassical limit,
more POs are necessary to explain the quantum dynamics of our system,
as discussed in Ref.~\cite{ver}.
\section{Summary and Conclusions}
\label{sec:final}

The introduction of Gutzwiller's trace formulae \cite{qchaos} in 1971
provided a method to understand the quantum mechanics of classically
chaotic Hamiltonian systems in terms of POs.
Its main drawback is the exponential growth in the number of orbits
involved as the Heisenberg time, $T_H$, increases, and many efforts
have been devoted to the development of resummation techniques that
improve its convergence.

Our work considers an alternative approach, initiated with
Refs.~\onlinecite{ver,PRE}, based on the use of special functions
well localized on a few POs and its manifolds,
the number of which grows only linearly with $T_H$.
To construct these localized functions we have used the method
described in Ref.~\onlinecite{pol}, based on the dynamics of a
coherent wave packet averaged along a given PO and its neighborhood.

In this paper we have studied in detail some important characteristics
and properties of these non--stationary scarred functions in the case
of the horizontal PO in the desymmetrized Bunimovich stadium billiard,
a paradigmatic example of chaos.
In particular, we have considered how the
localization properties, both in configuration and phase space,
of these functions calculated at energies, $E_0$, close to the
Bohr--Sommerfeld quantization conditions vary with the averaging
time, $T$. Our results show that the probability density in phase
space extends along the manifolds of the scarring PO, even when
the associated dynamics have left the linear regime.
In the second place, the transition to the semiclassical limit has
also been investigated by simultaneously changing $E_0$ and $T$.
As it takes place, all length features get better defined,
for example the focal points corresponding to the hyperbolic structure
around the fixed point.
Finally, we have also studied the dynamics of these non--stationary
wave functions.
When the corresponding probability density in phase space is followed
a dispersion along the manifolds is observed. In this process a
building up of probability peaks takes place on a few POs related
to the original one (the horizontal PO in our case).
The recurrences in the corresponding autocorrelation function can be
very important, and we have even shown one example in which they
are particularly pronounced.

To conclude, we think that this work provides valuable information that
can help in the disentanglement of the quantum mechanics of very chaotic
Hamiltonian systems in terms of simple classical structures.
\section*{Acknowledgments}

This work was partially supported by
PICT97 03--00050--01015, SECYT--ECOS (Argentina), and DGES (Spain)
under contracts no.\ PB96--76, PB98--115, and BMF2000--437.
DAW gratefully acknowledges support from CONICET (Argentina) and
AECI (Spain).
%

\end{multicols}

\begin{references}
\bibitem{footnote} Notice that the first homoclinic intersection of
  the stable and unstable manifolds of the horizontal PO also takes place
  in this region.
\end{references}
\end{document}